# Identification of error sources in high precision weight measurements of gyroscopes

I. Lőrincz and M. Tajmar*
*Institute of Aerospace Engineering*
*Technische Universität Dresden, 01307 Dresden, Germany*

**Abstract**

**A number of weight anomalies have been reported in the past with respect to gyroscopes. Much attention was gained from a paper in Physical Review Letters, when Japanese scientists announced that a gyroscope loses weight up to 0.005% when spinning only in the clockwise rotation with the gyroscope's axis in the vertical direction. Immediately afterwards, a number of other teams tried to replicate the effect, obtaining a null result. It was suggested that the reported effect by the Japanese was probably due to a vibration artifact, however, no final conclusion on the real cause has been obtained. We decided to build a dedicated high precision setup to test weight anomalies of spinning gyroscopes in various configurations. A number of error sources like precession and vibration and the nature of their influence on the measurements have been clearly identified, which led to the conclusive explanation of the conflicting reports. We found no anomaly within $\Delta m/m<2.6\times10^{-6}$ valid for both horizontal and vertical orientations.**

**Keywords**

Gyroscopes, weight anomalies, vibration artifacts, error sources

## 1. Introduction

A number of weight anomalies have been reported in the past with respect to gyroscopes, starting from the claims of Laithwaite in the 1970s, who designed a propellantless propulsion system using a special arrangement of spinning gyroscopes[1]. Much attention was gained from a paper in Physical Review Letters, when Japanese scientists announced that a gyroscope loses weight up to 0.005% when spinning only in clockwise rotation when viewed from above with the gyroscope's axis in the vertical direction[2]. In their setup, the gyroscope was encapsulated in a vacuum container measured on a counter-weight chemical balance. Immediately afterwards, a number of other teams tried to replicate the effect using a standard electronic balance all obtaining a null result[3,4,5,6,7]. It was suggested that the Hayasaka and Takeuchi effect was probably due to a vibration artifact[6,7], however, no final conclusion on the real cause has been obtained. On the other hand the data from Quinn and Picard[6] showed that small anomalies were present in some of the measurements which showed dependence on rotational speed. These authors suggested that vibration, friction torque and temperature drifts would be the cause for the measured anomaly. Also Dmitriev and Snegov reported a clear change of the mass of rotating bodies as a function of rotational speed[9]. The major cause for these anomalies was attributed to sensor drift and/or vibration, but in this case detailed analyses that would confirm these assumptions were absent.

A series of experiments has been designed in order to obtain a definite explanation of the conflicting reports. A summary of the most relevant previously published experimental results can be seen in Table 1.

Table 1: Summary of Relevant Published Experiments

| Experiment | $\eta = \lvert\delta M/M\rvert$ | Measurement Method | Orientations | Comments |
|---|---|---|---|---|
| Hayasaka-Takeuchi (1989)[2] | Up to $6.8 \times 10^{-5}$ | Counter-weight balance | Vertical | Anomalies measured |
| Faller et al. (1990)[4] | $< 9 \times 10^{-7}$ | Counter-weight balance | Vertical | No anomalies measured |
| Quinn-Picard (1990)[6] | $< 2 \times 10^{-7}$ | Flexture strip balance | Vertical | Anomalies present |
| Nitschke-Wilmarth (1990)[5] | $< 5 \times 10^{-7}$ | Electronic scale | Vertical North-South | No anomalies measured |
| Imanishi et al. (1991)[8] | $< 2.5 \times 10^{-6}$ | Electronic scale | Vertical | No anomalies measured |
| Dmitriev et al. (2001)[9] | Up to $2.8 \times 10^{-6}$ | Electronic scale | Counter-rotating gyroscopes Vertical/Horizontal | Anomalies present in horizontal orientation |

## 2. Description and Results

Our first setup reproduced the experiment of Hayasaka and Takeuchi[2] in which the measurement device was a counterbalance. The second reproduced the follow-up experiments in which the gyroscopes were measured directly on an electronic balance. In both experiments the same gyroscope was used as the test mass, which was a commercially available precision gyroscope [from Brightfusion Ltd, model name: “Super Precision Gyroscope”], comprised of a brass rotor with a mass of 112 g and a standard DC motor with which a maximum rotation speed in excess of 25000 RPM could be reached for relatively short periods (seconds) and approx. 19000 RPM for longer periods (minutes) [13000 RPM for Hayasaka and Takeuchi[2]].

Since in the reports of Hayasaka and Takeuchi [2] the results varied in function of rotational direction, we performed each measurement for six rotational vector orientations: up, down, north, south, east and west.

Here we will refer to the counter-balance layout as a balance and the electronic balance as a scale for convenience.

*a) Counter-Balance Experimental Set-up*

Figure 1 shows the schematic layout of the balance, where we can observe the solutions for the elimination of first order error sources. They comprise a sealed container which would restrict any airflow from within the measurement system to the outside environment and a wireless data interface. In all of the previously published experiments a sealed container was also considered a necessity in order to restrict airflow; in some, the container was also evacuated in order to minimize the effects of airflow. We decided that a sealed environment would be sufficient without evacuation.

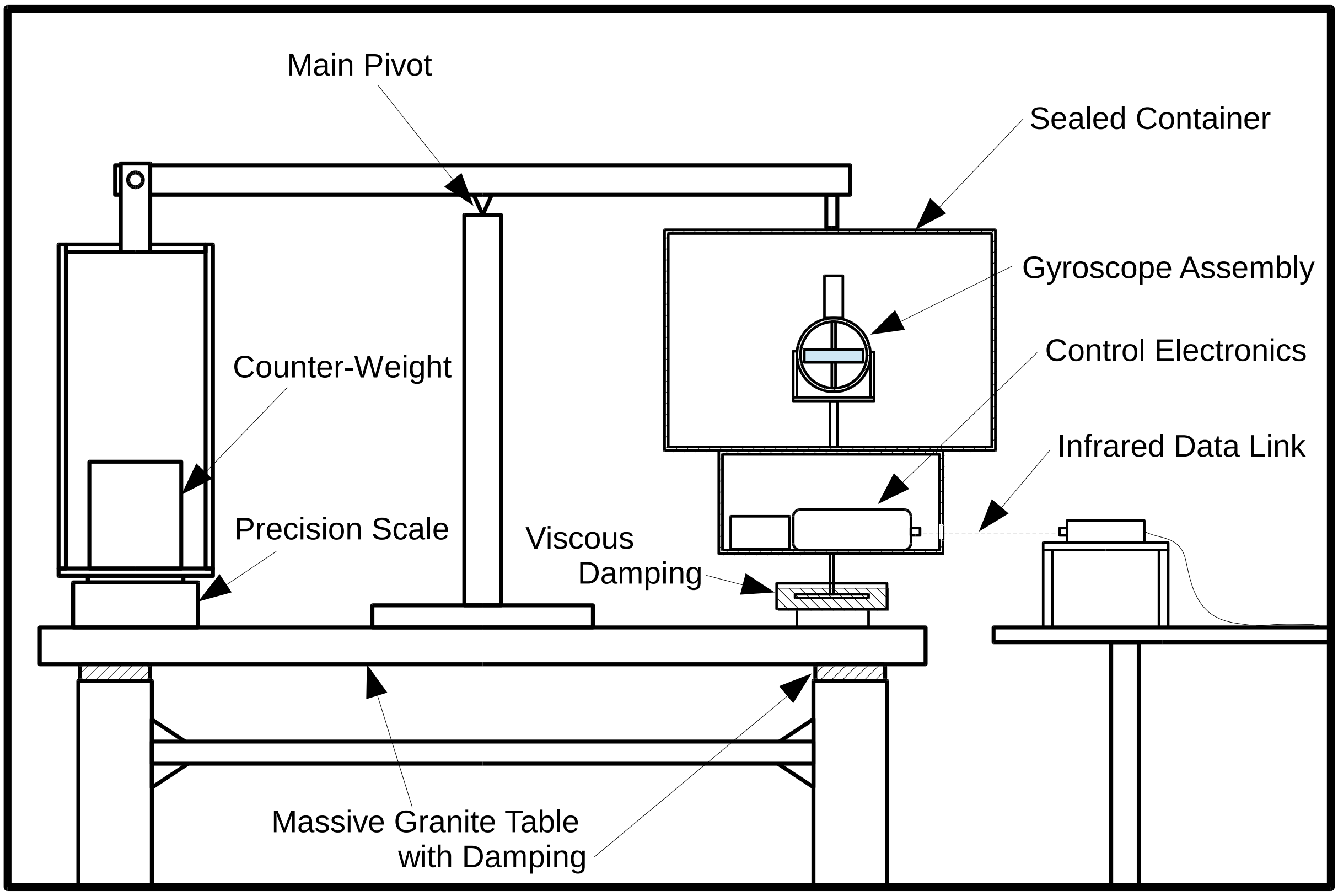

**Figure 1: Schematic Setup using Balance Arm and Counter-Weights (Counter-Balance Setup)**

In previous experiments wires were used for the control of the gyroscopes but we considered that this solution introduces an error source; hence we implemented a wireless control interface with the power supply (batteries) and control electronics for the gyroscope all placed inside the container. For the actual weight measurements we applied an electronic precision scale [Sartorius AX 224] with a maximum range of 220 g and a 0.1 mg resolution that was placed under the counter-weight holder. The output of the scale is a selectable unit of mass (grams in our case) hence the actual internal algorithm through which the device computes the mass from the force applied on top of the measuring plate is unknown.

We observed that the scale was also influenced by the position of the contact point between the object to be measured and the scale's plate. In order to eliminate this possible error source that would originate from the surface imperfections of two larger plates (undefined contact points), the physical connection between the counter-weight holder and the scale was realized through a solid-spike (generally used for hi-fi speaker support), which gave a well-defined small contact area. The solid-spike has the shape of a cone that is made out of hardened steel with a thread on its base (see Figure 2). The apex of the cone has a small radius of approx. 0.4 mm, thus contact between the spike and the plate is reduced to a small point. This solution allowed the fine tuning of the weight load on the scale and also the adjustment of the horizontal angle of the main balance arm. A digital level was placed on top of the balance arm and by using the spike's thread the angle could be set with a precision of ±0.05 degrees.

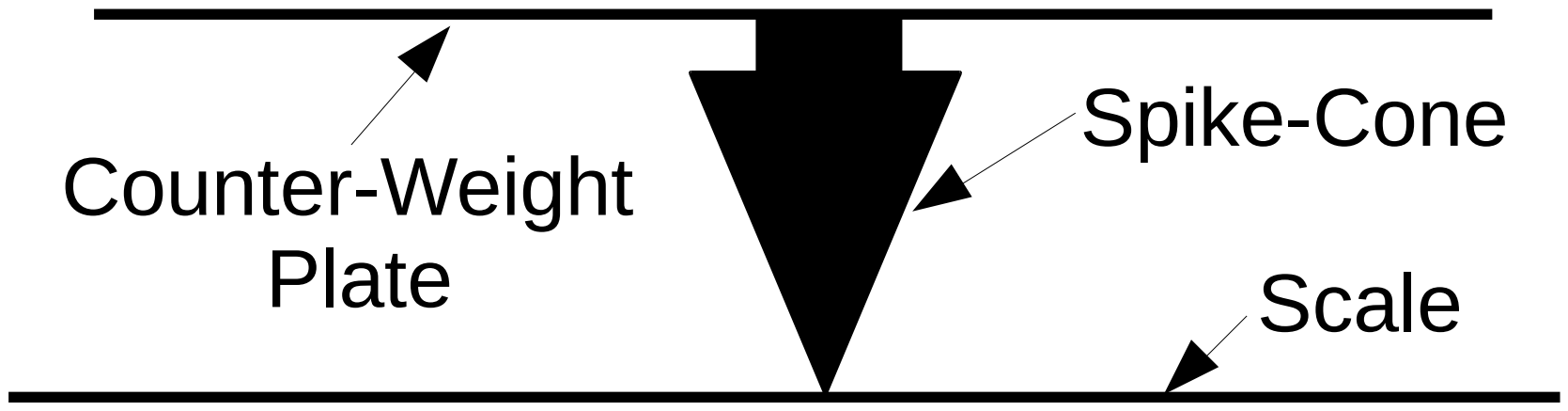

**Figure 2: The schematic of the solid-spike assembly**

The central pivot point was chosen to be at the center of the distance between the suspension points of the container and counter-weight holder, dividing the balance arm into two equal lever lengths of 330 mm ± 0.01 mm, resulting in a 1:1 lever ratio.

The next step was to identify and minimize higher order error sources. Environmental vibrations would fall into this category, which were minimized through a damped massive granite table (see Figure 1), specially designed for high precision weight measurements. By placing the whole setup in an airtight room all the environmental error sources were minimized to the best of our abilities. The experiment room was sealed from the outside and the operation of the electronics was carried out from another control room. Since even the cooler from the control PC could influence the measurement (induced airflow) extra care was taken to isolate it from the experiment by closing its openings and further enclosing it in a container.

A test was carried out to see if induced electromagnetic fields would influence the scale by putting a small calibration weight on the scale and lifting the spike from the scale. This way the mechanical contact was severed between the scale and balance so if a weight change would have appeared it would have been caused by unwanted influences (most probably electromagnetic radiation). The test confirmed that no influences were present during the measurements. Thus the environmental parameters of the room (temperature, electromagnetic fields, air pressure, etc.) could be regarded as constant during a measurement.

The next step was to analyze the unwanted influences which originated from the inside of the experiment's system. The complex structure of the balance resulted in a large number of resonance modes at higher frequencies. After we implemented the viscous fluid damping solution, the resonance effects were minimized below the threshold of our measurement precision. This was determined after the sudden weight changes disappeared during the acceleration or deceleration of the gyroscope.

Two critical points could be identified on the balance: the central pivot of the balance arm and the suspension of the sealed container. The suspension of the counterweight holder was implemented by using two ball bearings which made the connection stiff except for the rotation axis (parallel to the rotation axis of the central pivot). The load on this pivot was set by the load on the scale (in the order of grams) and compared to the load on the central pivot (in the order of kilograms) it would be only interesting for a higher order analysis, which would lie beyond our measurement's precision.

Since the angular acceleration of the gyroscope's rotor was present within the system, the production of the resulting torques is an inevitable consequence and thus it had to be treated as an error source. Even in the case when the speed was to be held constant during a measurement unwanted torques would be present due to variations of friction within the gyroscope's bearings, unwanted drifts and sudden noise peaks in the control voltage of the DC motor brought about by temperature, mechanical load variations, etc.

In all cases in which rotation is present in a measurement system, the possibility of precession has to be carefully analyzed, since the effect is in most cases counterintuitive or even unexpected. The effect of forced precession on weight change was investigated by Wayte[10], who could achieve an apparent mass reduction of up to 8% depending on rotation speed and the relative measurement point within the system (Figure 3).

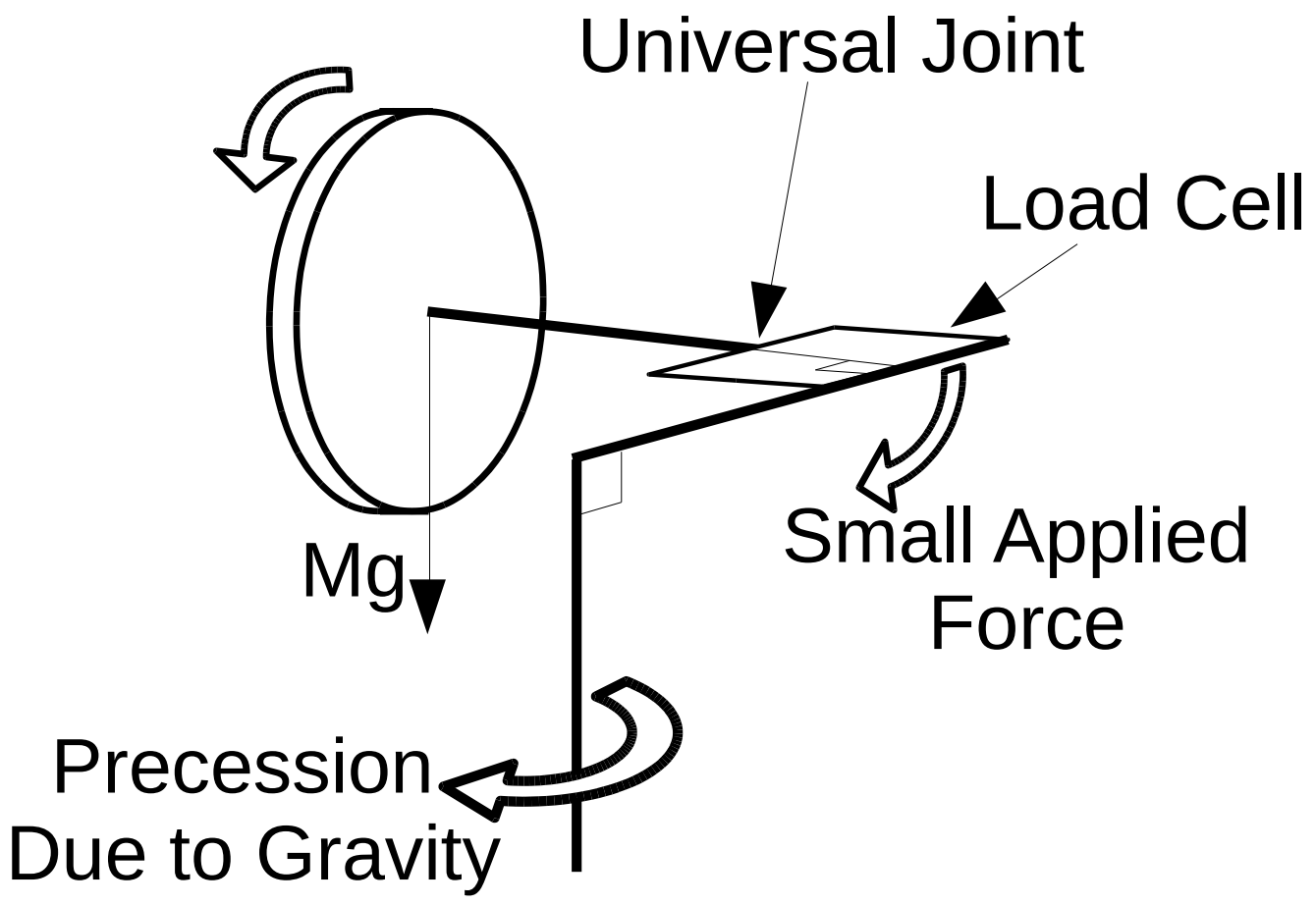

**Figure 3: Schematic of Wayte's[10] forced precession experiment**

As a first step to exclude the possibility of precession we used two solid-spikes for the central pivot, which provided a very low friction pivot with a two point suspension, thus eliminating the possibility of rotation in the horizontal plane. The spikes were aligned transverse (90 degrees) to the longitudinal axis of the balance beam and protruded into conical cups with smaller apex angles cut into stainless steel support pillars (Figure 4). Thus the contact between the spikes and the pillars were circles, which allowed no horizontal slippage while they were under the normal load of the whole assembly. The cups were filled with high quality conductive oil for a good electrical contact between the grounded pillars and the rest of the assembly (to assure the same electrostatic potential) and to minimize friction.

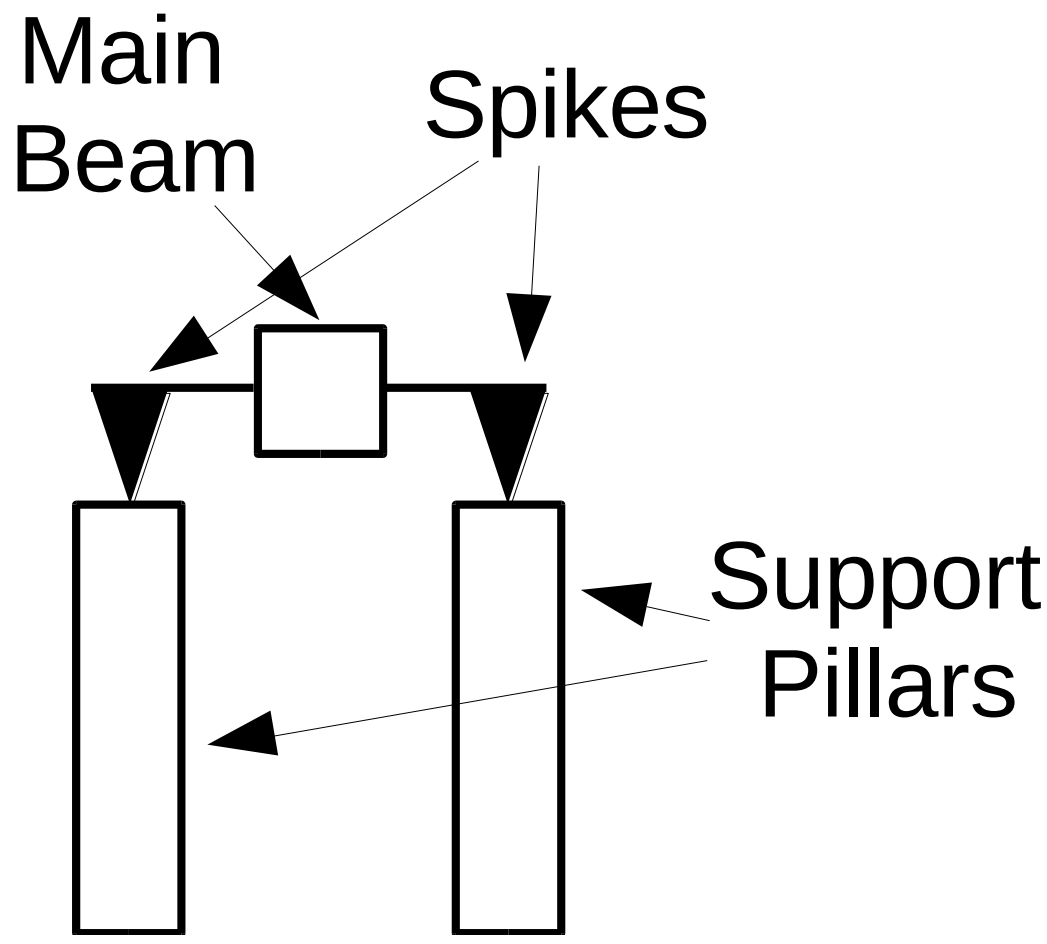

**Figure 4: Main pivot solid-spikes arrangement**

Two possible solutions could be identified with respect to the torque and precession problem:

- The suspension of the container should not allow rotation around any axis (a stiff connection)

- The suspension should allow unrestricted rotation (pivoted connection – acting as a hinge around every axis of rotation).

The first option would eliminate the possibility of precession (assuming that the only rotation source is the suspension point of the container), but the generated torque would be directly transferred to the main pivot point. The second option would not allow the transmission of the produced torque to the main pivot but precession of the sealed gyroscope container would be unrestricted. One of these cases had to be present in the Hayasaka experiment, even if these details were not described, hence we tested both situations.

Numerous measurements were taken with the balance setup while the gyroscope's orientation and rotation speed were varied. The qualitative assessment of the measurements can be summarized in two figures. In Figure 5 the case with the stiff connection between the container and the balance arm is shown where the gyroscope's spin axis orientation was coincident with the orientation of the balance arm's rotational axis.

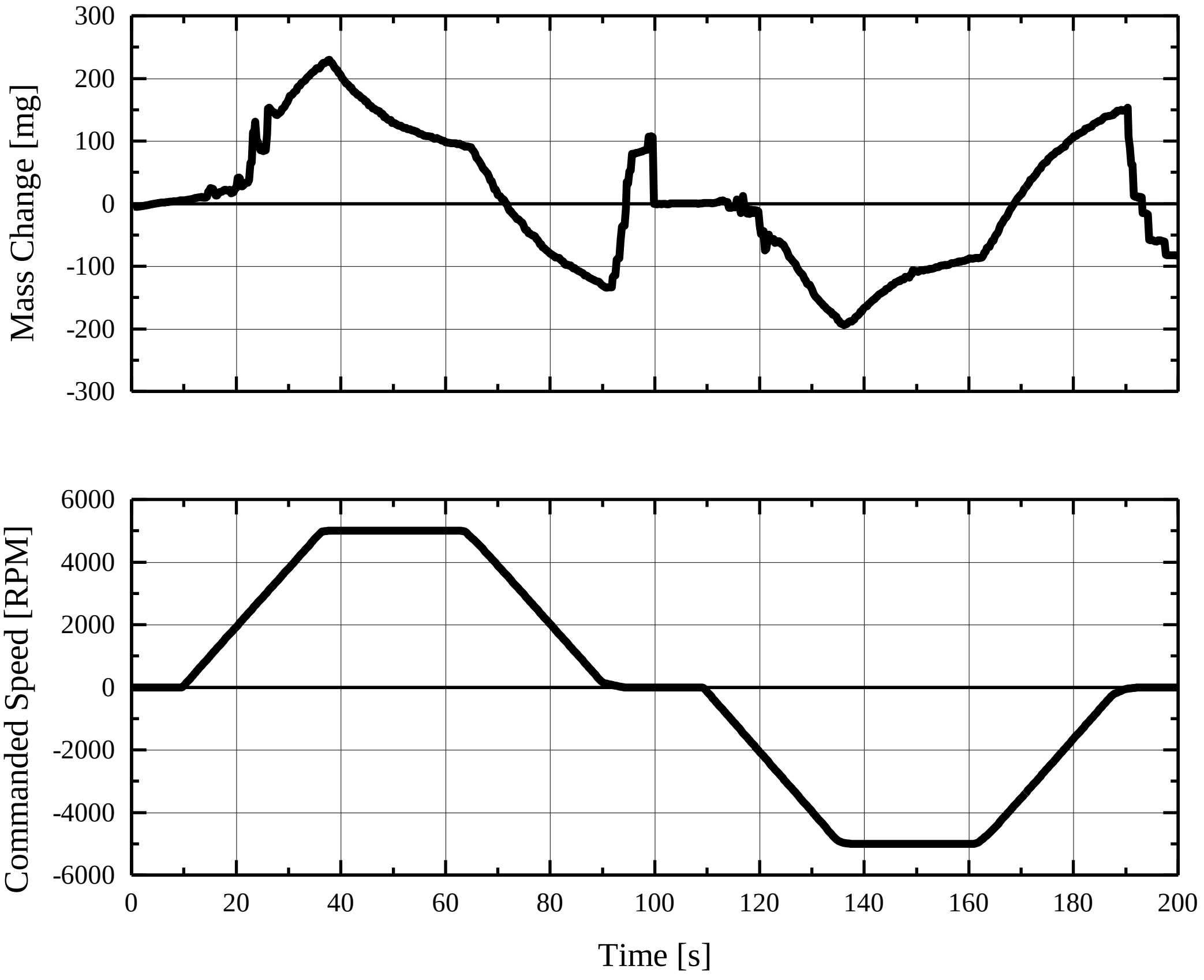

**Figure 5: Measurement of Gyroscope Mass Change versus Commanded Speed in Counter-Balance Setup with Vertical Spin Plane inside Coupled Container (Stiff connection)**

In this situation the influence of torque on the mass measurement is at maximum, as we can see the maximum apparent mass change is in excess of 200 mg at the maximum acceleration of the gyroscope. The apparent mass change due to the torque can be easily calculated if the torque is decomposed into a force couple, where the couple components act at the suspension points on the main balance arm (see Figure 6).

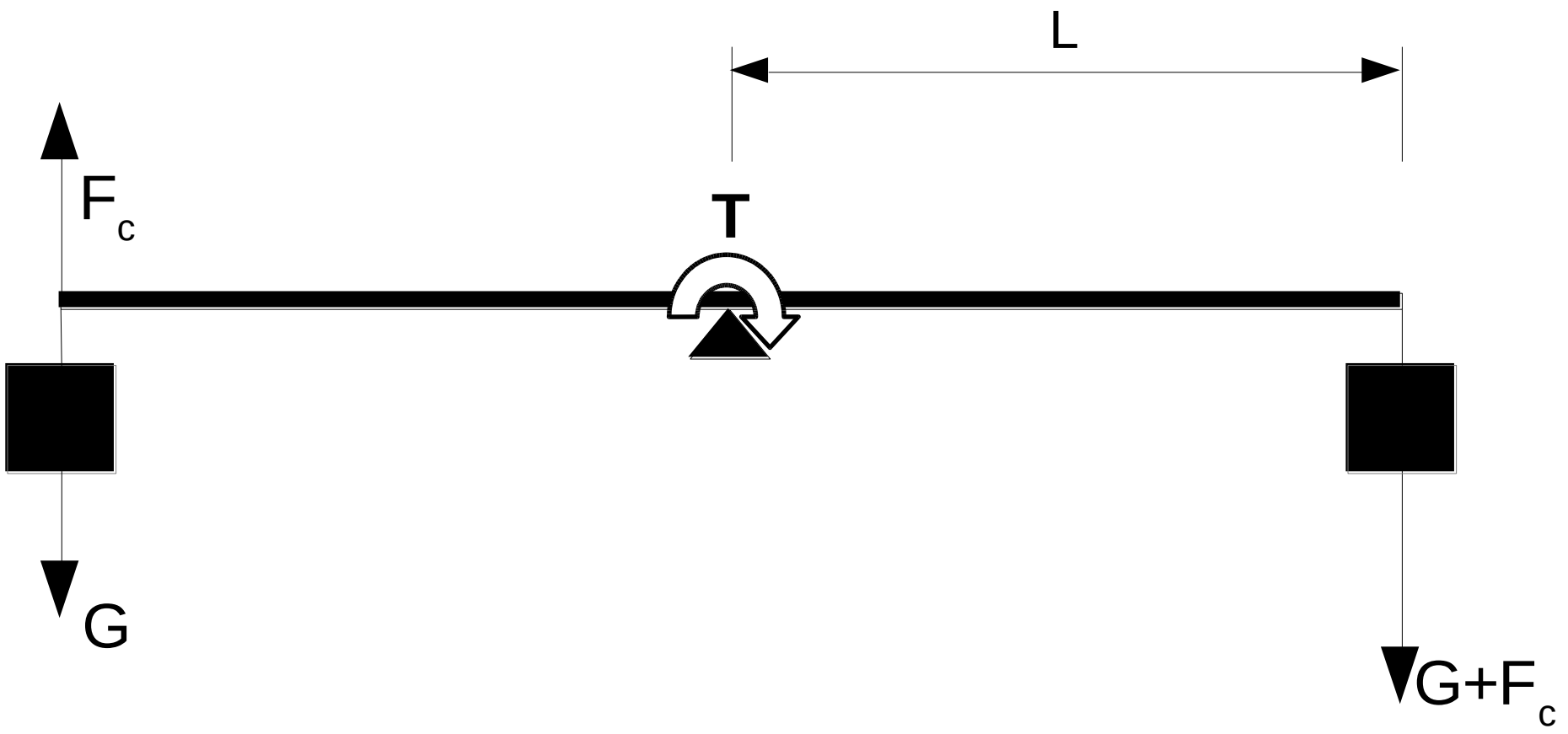

**Figure 6: Torque (T) decomposition into a force couple ($F_c$) on a counter-weight balance, where G represents the real weight**

We can write:

$$F_c = m_c g_0 \tag{1}$$

where $g_0$ is the gravitational acceleration and $m_c$ is the apparent mass, which causes the force couple component $F_c$ in the gravitational field. In this case the weight is measured on only one side of the balance, meaning that both of the force couple's components will be causing a weight change in the same direction (e.g.: on one side making the counter-weight appear lighter, on the other side making the container appear heavier). The measured weight change will become:

$$\Delta M = 2F_c = 2m_c g_0 \tag{2}$$

where $\Delta M$ is the apparent weight change. According to the definition of the force couple we can write:

$$T = Fc \cdot L \tag{3}$$

where T is the torque generated by the gyroscope assembly and L is the half-arm length of the balance. Thus the apparent mass change is given by:

$$\Delta m = 2m_c = 2\frac{F_c}{g_0} = 2\frac{T}{Lg_0} \tag{4}$$

where $\Delta m$ is the measured apparent mass change from the quiescent state to the situation where a torque is acting on the main pivot. We can see that even a small torque can produce significant measurement accuracy issues. In our case, where the half-arm length is 330 mm, a mass change of 200 mg (see Figure 5) is caused by a torque of approx. 0.356 mNm. This effect can be minimized if the gyroscope's rotation vector is aligned with the vertical axis. Since perfect alignment is unrealistic, we can easily calculate the necessary minimal angular misalignment with respect to the vertical in order to obtain a smaller apparent mass change than the measurement's resolution (0.1 mg in our case). If we assume the same value for the torque as previously obtained, we get:

$$\operatorname{asin}\left(\frac{\Delta m L g_0}{2T}\right) = 0.026° \tag{5}$$

Thus we can conclude that it would be highly unlikely to achieve the necessary orientation precision with standard tools (that would be available within small scale projects), which would assure a smaller torque influence than the measurement's resolution. A realistic value for the orientation precision would be a few degrees that would correspond to a torque about two orders of magnitude smaller than our assumed value.

In Figure 7 the pivoted connection between container and balance arm is applied while the orientation of the gyroscope's spin axis is vertical.

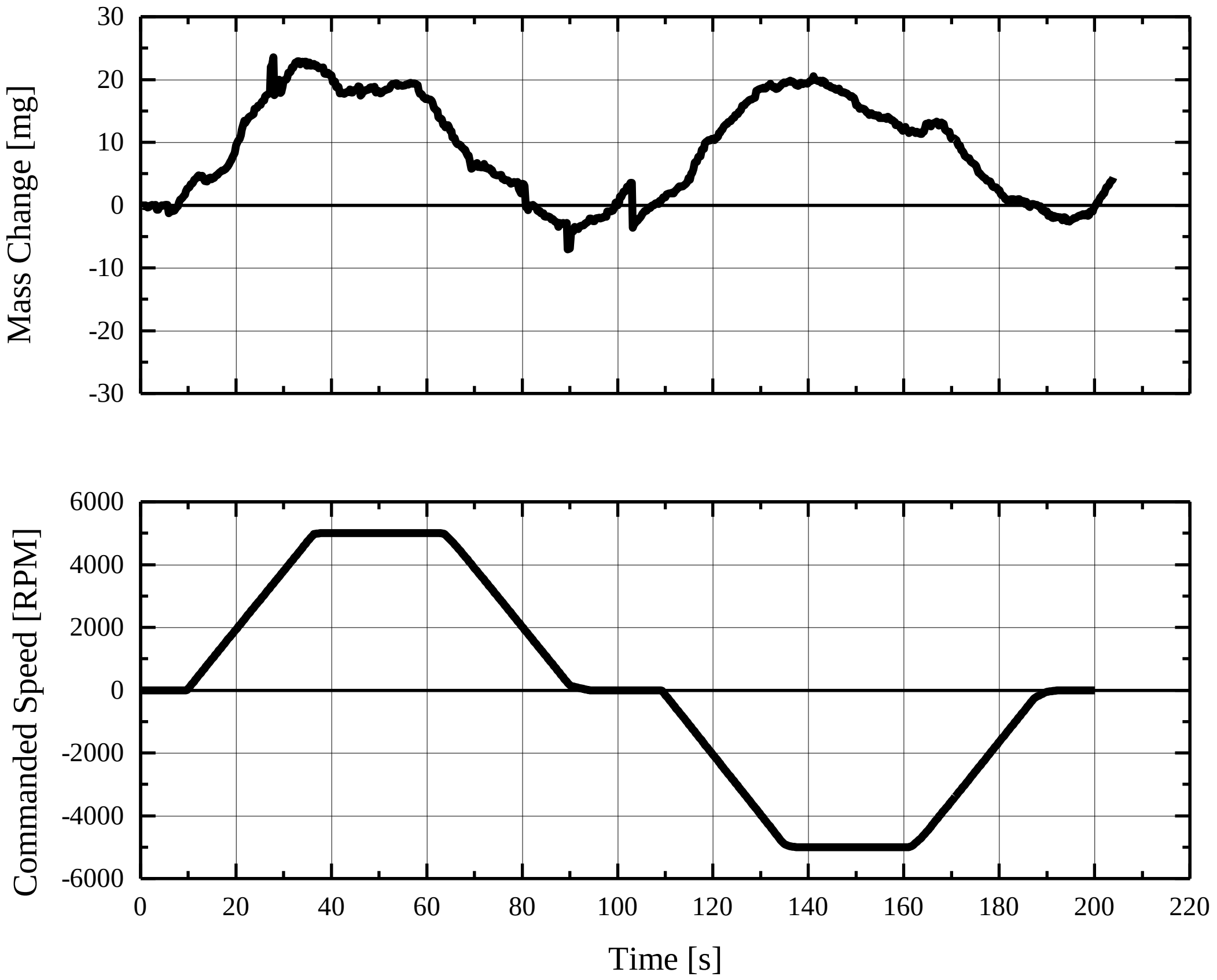

**Figure 7: Measurement of Gyroscope Mass Change versus Commanded Speed in Counter-Balance Setup with Horizontal spin plane inside uncoupled container (Pivoted connection)**

In this case we can see a reduced apparent mass change by one order of magnitude, which is systematically (independent of rotation direction) caused by precession. These cases serve as the baseline for the quality assessment of the balance setup. The same reasoning regarding orientation tolerance, as for the first case, can be applied for the second option, the pivoted connection. Thus the conclusion has to be drawn that the counterbalance method for the weight measurement of gyroscopes is unreliable and should be avoided.

The next error source from within the setup, which had to be attended to, was the vibration caused by the gyroscope and the driving motor. Since these could not be directly influenced we had to implement a special damping solution that would act on the whole balance. It is evidently crucial that the damping should not influence the steady state mass measurements, which would be the case with a solution based on dry friction (static friction). Another aspect, which is not that evident, is that the application of an active damping system would create an unidentifiable error source due to the unknown vibration induced by the scale's feedback system. The scale has also a feedback system to compensate for vibrating loads, which is unknown to the customer. Hence the combination of an unknown active vibration compensation (which is effectively an independent vibration according to our observations) and an active damping system would result in an unknown vibration profile with possible transient resonance frequencies. Our solution was to apply viscous friction, by immersing a flat plate, connected to the balance, into a viscous fluid (see Figure 1). With this solution we could make sure that the steady state mass measurement was not influenced and we could also tune the system to reach critical damping by changing the viscosity of the fluid. We used a common commercially available viscous fluid that was soluble in water.

*b) Electronic Scale Experimental Set-up*

For the second experiment type, a small airtight container was made out of transparent PMMA (see Figure 8) in which the gyroscope's frame was tightly fixed with screws to the container walls. The procedure that we applied is the same as the one in the Quinn and Picard[6] experiment where the rotor was spun up and then placed directly on the scale while it slowly decelerated. We also used a laser RPM sensor to accurately keep track of the rotation speed during the measurements. This situation was a lot simpler with respect to the identification of error sources. Since the airflow was restricted within the PMMA container and no relative displacement was allowed as it was glued the scale's plate (it has to be noted that the damping between container and scale as shown in Figure 8 was not always applied), we only had to analyze the vibrations and their effect on the measurements. In this scenario two vibration sources were present: one generated by the gyroscope and another generated by the scale's active measurement system. The only way to influence the interaction of these two sources was to place various damping materials in-between the two.

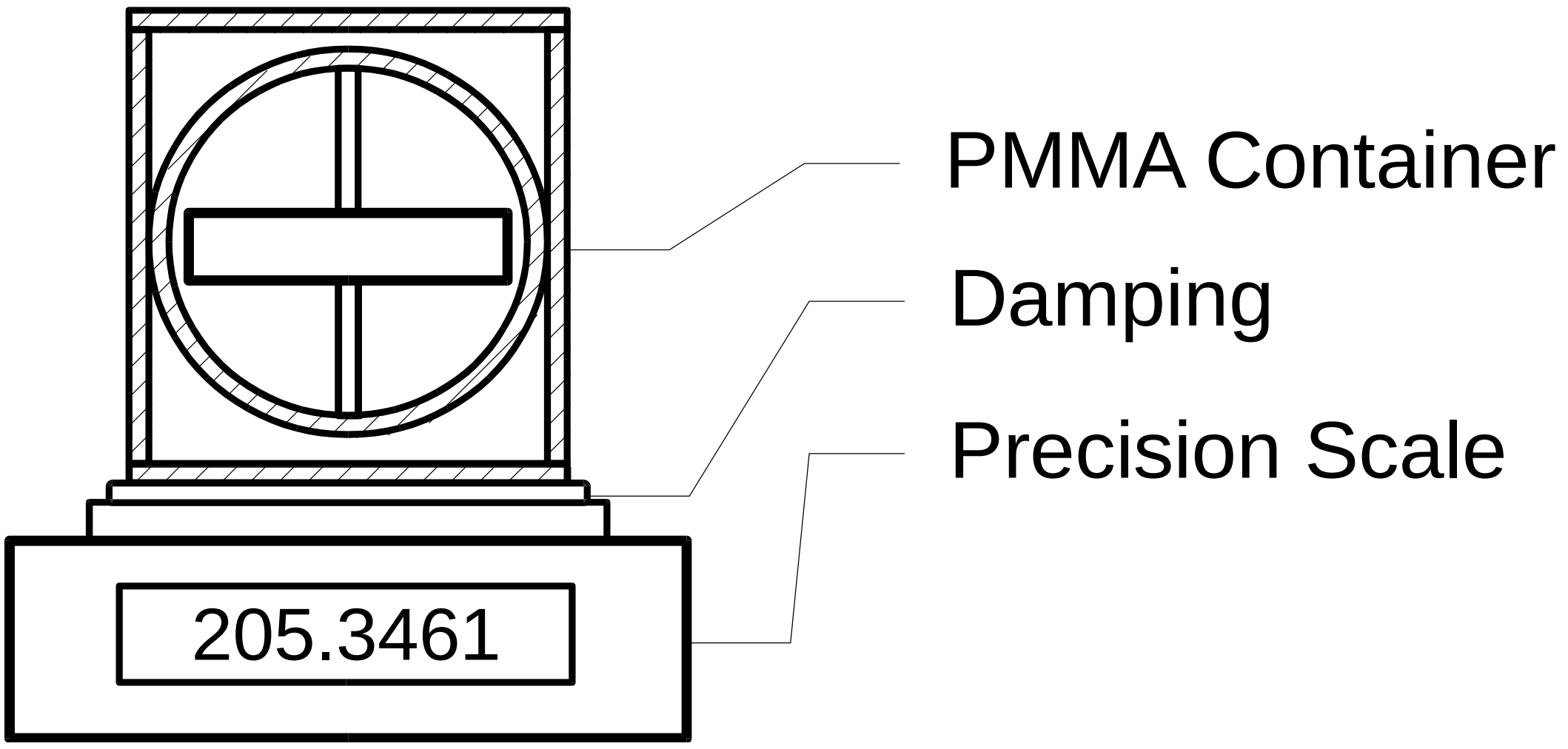

**Figure 8: Schematic Setup with Gyroscope Directly on the Scale (Electronic Scale Set-up)**

In Figure 9 we present the results without damping and without the container being glued to the scale. In both cases the measurements show similar mass anomalies as in the cases of Quinn and Picard[6] and Dmitriev and Snegov[9], which seem to be dependent on rotation speed and direction. No satisfactory conclusive explanation was given by others to these anomalies.

In all of our measurements we could identify small regions where peaks of mass change appeared as a function of rotation speed. These regions varied (amplitude and RPM interval) if the mechanical coupling between scale and container was changed, suggesting that these were resonance zones generated by the interaction of the two vibration sources. Based on this assumption we tried to eliminate or minimize the resonance amplitudes by mechanically decoupling the vibration sources.

In Figure 10(a) a measurement result is shown in which the container was glued to the scale. The measurement plot changed from an apparent smooth function to a discontinuous one. Here we can clearly identify three regions where an apparent mass change is dominant, suggesting that resonances appeared around 9000, 12000 and between 16000 to 19000 RPM. Secondary resonance zones were also present in some cases which can be also identified in the reported measurements of Quinn and Picard[6].

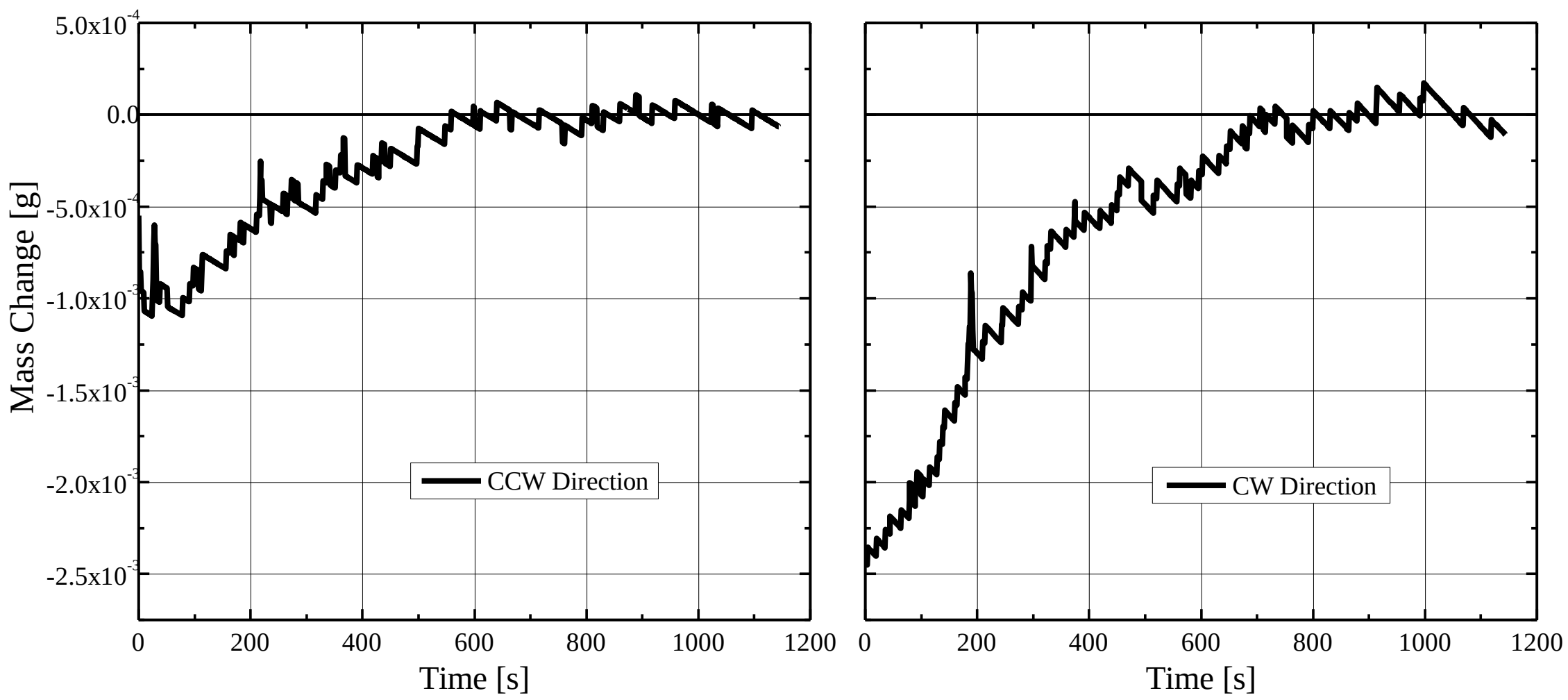

a.) Initial Gyro Speed = 13600 RPM

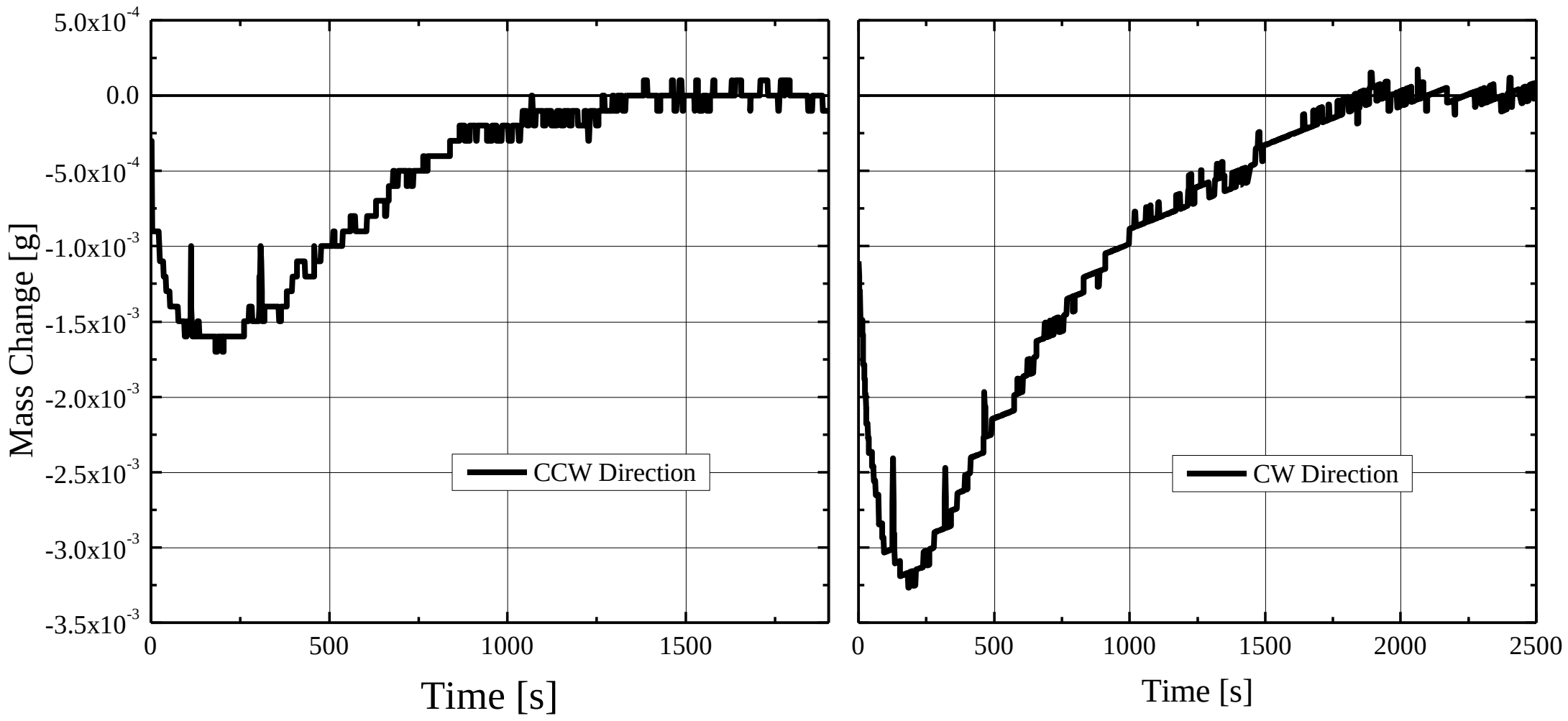

b.) Initial Gyro Speed = 25400 RPM

**Figure 9: Measurement of Gyroscope Mass Change versus Time - Scale Without Damping, Gyroscope Not Fixed to Scale**

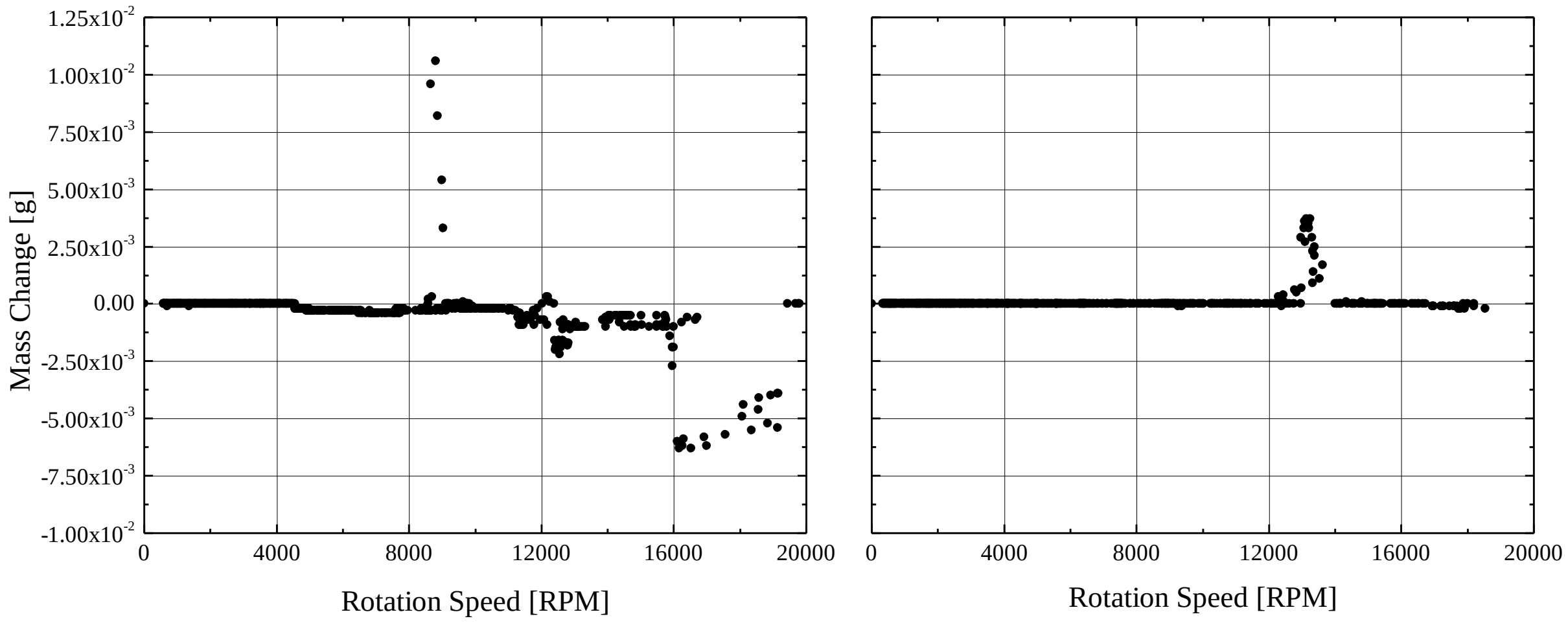

a.) Without Damping

b.) Properly Damped Gyroscope

**Figure 10: Measurement of Gyroscope Mass Change versus Rotation Speed – Gyroscope Fixed to Scale**

After testing more than 12 damping materials (mostly various types of foams) between the gyroscope and scale we found the proper damping material with which the amplitude of the resonance zones was minimized (see Figure 9(b)) to the point where a linear fit analysis could be reliably performed.

In Table 2 the summary of the linear fit/regression analysis of the measurements for every direction and orientation is presented both for undamped and damped cases.

Table 2: Regression Analysis of the Mass Changes versus Rotation Speed (up to 20000 RPM) for the Gyroscope fixed to Scale with and without Damping

| | **Undamped*** | | **Damped*** | |
|---|---|---|---|---|
| **Orientation/ Direction** | **Slope [g/RPM]** | **Standard Error** | **Slope [g/RPM]** | **Standard Error** |
| **Vertical/CW** | $3.1 \times 10^{-8}$ | $1.3 \times 10^{-8}$ | $4.0 \times 10^{-9}$ | $1.9 \times 10^{-9}$ |
| **Vertical/CCW** | $-6.4 \times 10^{-8}$ | $2.9 \times 10^{-9}$ | $-6.8 \times 10^{-9}$ | $2.0 \times 10^{-9}$ |
| **North-South/CW** | $-1.0 \times 10^{-7}$ | $1.7 \times 10^{-9}$ | $-1.3 \times 10^{-9}$ | $1.8 \times 10^{-9}$ |
| **North-South /CCW** | $-7.2 \times 10^{-8}$ | $7.4 \times 10^{-9}$ | $1.8 \times 10^{-9}$ | $3.7 \times 10^{-10}$ |
| **West-East/CW** | $2.1 \times 10^{-8}$ | $5.3 \times 10^{-9}$ | $-1.8 \times 10^{-9}$ | $1.4 \times 10^{-9}$ |
| **West-East /CCW** | $-3.1 \times 10^{-8}$ | $5.8 \times 10^{-9}$ | $9.2 \times 10^{-9}$ | $7.7 \times 10^{-10}$ |

*Balance Resolution $\pm 5 \times 10^{-9}$ g/RPM

*c) Numerical Simulation of the Observed Resonances*

In order to verify that indeed such resonances could be the cause of the measurement anomalies, a numerical simulation was performed of a simplified but similar case. As the baseline for the simulation we assumed that an electronic scale obtains the mass of a test body by calculating the force necessary to counteract the gravitational force acting on the body by some sensor-actuator assembly. In some cases no actuators are present in the assembly, but since it could be observed that the scale could act as an active vibration source it can be clearly stated that an actuator was applied in our scale. Since the mechatronics of such a system has to be based on a feedback loop, the actual measurements must be acquired at a specific frequency. Depending on the design and the user-setup of the scale's mechatronics this frequency can vary in time. Our first simplifying assumption was to consider a scale with a constant measuring frequency. Since this frequency can be relatively high, it is logical to assume that an averaging algorithm is also included in the scale's software. We approximate this algorithm with the combination of a simple average over a specified short time period with a moving average.

The next important factor was the deceleration of the gyroscope, which would provide a second mechanical frequency that is varying in time. To accurately calculate this, the friction torque components had to be deduced from the measurements and together with the rotor's moment of inertia they give the accurate velocity profile:

$$\tau_{tot} = \gamma_v \omega + \tau_c \quad (6)$$

$$\ddot{\theta} = \frac{\tau_{tot}}{I_{rotor}} \quad (7)$$

where $\tau_{tot}$ is the total friction torque, $\gamma_v$ the viscous friction torque coefficient (kinetic friction torque), $\tau_c$ the coulomb friction torque, $\ddot{\theta}$ the angular acceleration and $I_{rotor}$ is the gyroscope's moment of inertia. The imbalance of the rotor, which would provide a first order approximation of the gyroscope's vibration, is simulated by a small point mass attached to the edge of the rotor. If we consider that the gyroscope's spin axis is horizontal, the component of the resulting disturbing force normal to the surface of the scale is given by:

$$F = m_u R \omega^2 \cos(\theta) \quad (8)$$

where $m_u$ is the small point mass, $R$ is the radius of the rotor, $\omega$ is the angular velocity and $\theta$ is the angle between the vertical axis of the coordinate system and the point mass. Thus in order to see the effect of a vibrating mass on such an electronic scale, we have to calculate the value of the normal component of the disturbing force with the specified constant frequency, while applying the averaging algorithm.

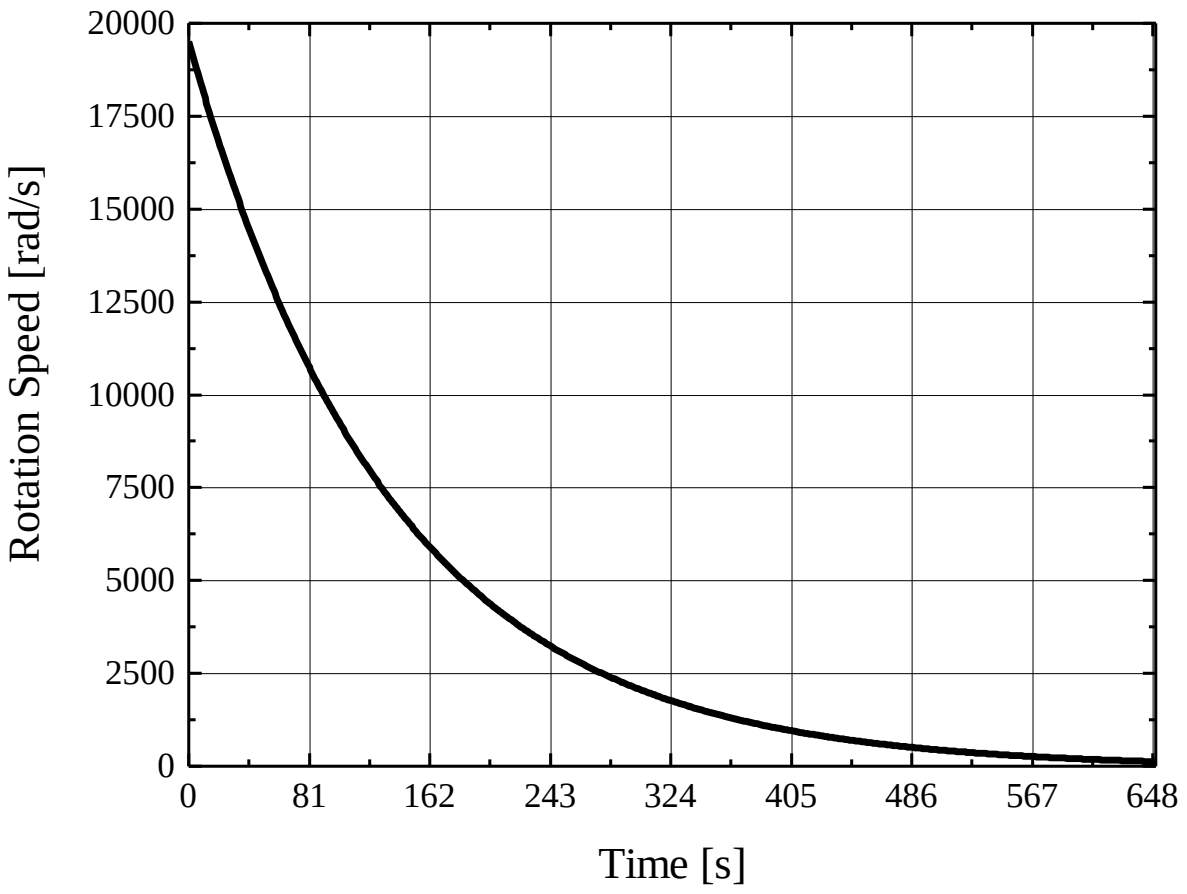

a.) Decay of Rotation Speed

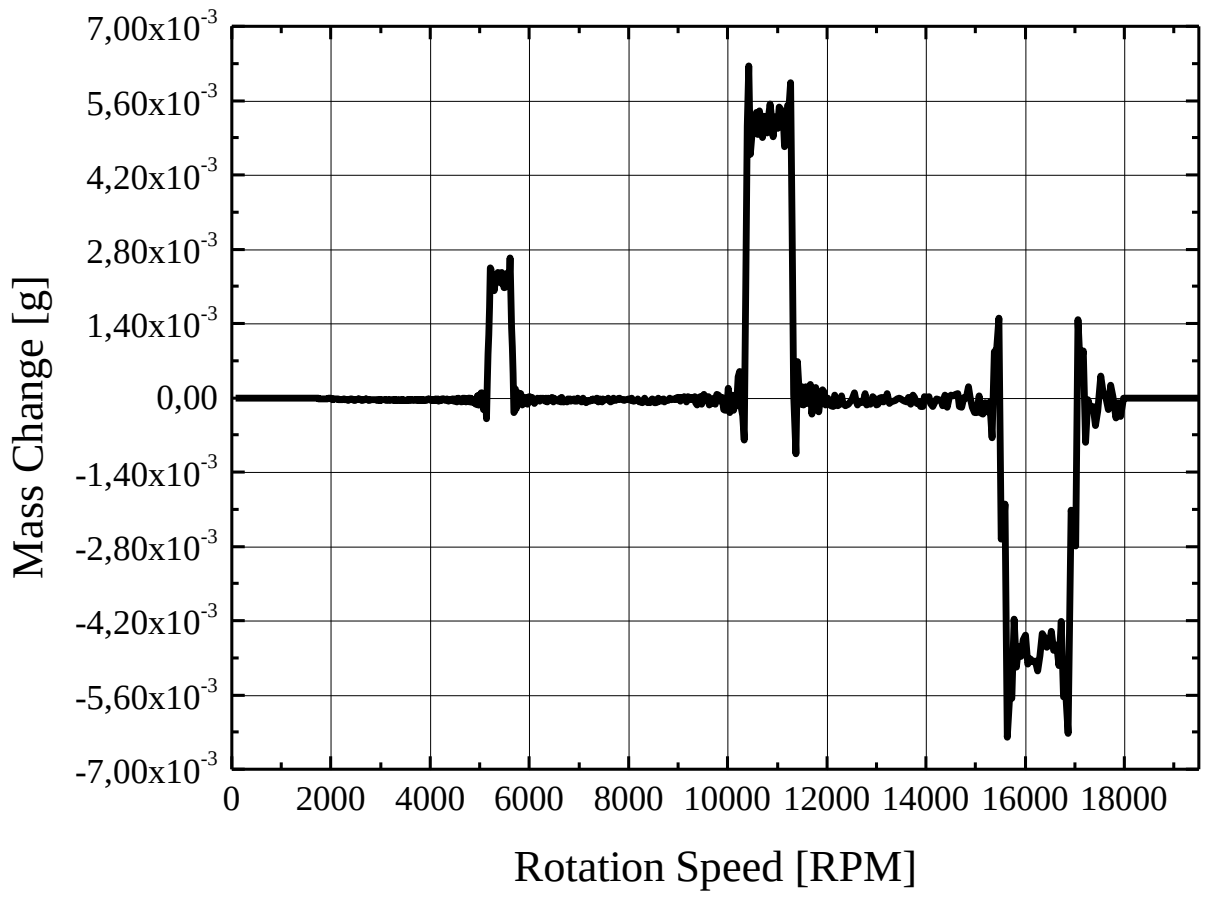

b.) Mass Change due to Vibration

**Figure 8: Simulation of a Rotating Vibration Source on an Electronic Scale**

The simulation result is presented in Figure 11, in which we assumed a scale frequency of 94 Hz and an unbalancing mass of 0.025 mg. We can clearly identify three resonance zones at approximately the same RPM bands as in Figure 10(a). In reality this is a much more complex interaction since the scale operates with an active feedback loop (producing force) that generates additional vibrations, which has a varying frequency and also an unknown averaging algorithm, but the principle remains valid. We could also observe that the magnitudes of the resonance zones are highly sensitive to changes in the measuring frequency.

## 3. Discussions and Conclusion

We reproduced two experiments to measure possible mass changes of spinning gyroscopes, one where the gyroscope was mounted on a counterbalance and another one with the gyroscope mounted directly on an electronic scale. Our goal was to find out why some groups did and others did not report anomalous weight changes. The obtained precision for the two experiments differed significantly and the counterbalance setup yielded the poorest performance. We demonstrated that a counter-weight balance is inadequate for the weight measurement of rotating bodies. We omitted to analyze the solution where two gyroscopes would rotate in opposite directions[9], which, in a first order approximation, would result in canceling out the unwanted precession torque and the torque generated by the rotor's angular accelerations. Although this approach would be an attractive solution, it would only increase the system's complexity without providing the necessary performance improvement, since a serious accuracy issue would be the synchronization of the two rotation speeds, with two motors or with a single motor and gears. A possible higher order error source in this situation would be the difference in the moment of inertia of the two rotors, leading also to apparent weight anomalies.

More importantly we have demonstrated that periodical vibrations, in our case generated by a spinning body, can significantly influence the readout of the scale, which probably operates based on an active feedback loop. The majority of electronic scales would fall into this category, since they use an electromagnetic feedback loop in order to determine the force necessary to counter the weight of a test mass. This feedback loop has a defined frequency (which is mostly a trade secret of the production companies) for any specific state. Thus the interaction of these two frequencies can lead to measurement errors. In case this situation cannot be avoided in a measurement, we advise that proper precautions should be taken in order to decouple the vibration sources. Finally we conclude that the reason for the conflicting reports of the mass measurements of spinning gyroscopes was due to the error sources presented in this paper.

Since the standard error for the damped cases is actually below the balance resolution (0.1 mg at 20.000 RPM giving $\pm 5\times 10^{-9}$ g/RPM), and since all slope values are below $3\sigma$ of this balance resolution standard error, we can conclude that all our measurements indicate a null result for possible mass change of gyroscopes within our balance resolution. Hence we confirm the null results obtained by the previous groups[2,4,5,6,8,9]. We get a relative mass change of:

$$\eta = \frac{\delta m}{m} < 2.6 \times 10^{-6}$$

where $\delta m$ is the mass change and $m$ is the rotor's mass. This result is similar to the one from Imanishi[8] – but valid for all gyroscope orientations along the horizontal and vertical axis.